\documentclass[12pt]{article}
\newlength{\extraspace}
\newlength{\extraspaces}
\catcode`\@=11
\def\numberbysection{\@addtoreset{equation}{section}
\def\theequation{\arabic{section}.\arabic{equation}}}
\newcommand{\newsection}[1]{
\vspace{7mm}
\pagebreak[3]
\addtocounter{section}{1}
\begin{center}
{\large {\bf \thesection. #1}}
\end{center}
\nopagebreak
\medskip
\nopagebreak
\hspace{3mm}}
\newcommand{\nonu}{\nonumber \\[.5mm]}
\newcommand{\A}{&\!\!\!}

\begin{document}
\addtolength{\baselineskip}{.7mm}
\thispagestyle{empty}
\begin{flushright}
STUPP--06--187 \\
\texttt{hep-th/0609119} \\ 
September, 2006
\end{flushright}
\vspace{20mm}
\begin{center}
{\Large \textbf{PSU($2,2|4$) Transformations of IIB Superstring \\[2mm]
in AdS${}_5$ $\times$ S${}^5$
}} \\[20mm]
\textsc{Madoka Nishimura}${}^{\rm a}$
\hspace{1mm} and \hspace{2mm}
\textsc{Yoshiaki Tanii}${}^{\rm b}$\footnote{
\tt e-mail: tanii@phy.saitama-u.ac.jp} \\[7mm]
${}^{\rm a}$\textit{Department of Community Service and Science \\
Tohoku University of Community Service and Science \\
Iimoriyama 3--5--1, Sakata 998--8580, Japan} \\[5mm]
${}^{\rm b}$\textit{Division of Material Science \\ 
Graduate School of Science and Engineering \\
Saitama University, Saitama 338-8570, Japan} \\[20mm]
\textbf{Abstract}\\[7mm]
{\parbox{13cm}{\hspace{5mm}
The PSU($2,2|4$) transformation laws of the IIB superstring theory 
in the AdS${}_5$ $\times$ S${}^5$ background are explicitly 
obtained for the light-cone gauge in the Green-Schwarz formalism.  
}}
\end{center}
\vfill
\newpage
\setcounter{section}{0}
\setcounter{equation}{0}
%
%
\newsection{Introduction}
The construction and the quantization of the superstring theory 
in anti de Sitter (AdS) spacetime have been an important subject 
since the original AdS/CFT correspondence 
\cite{Maldacena:1997re,Gubser:1998bc,Witten:1998qj} was proposed. 
Metsaev and Tseytlin \cite{Metsaev:1998it} constructed 
the Green-Schwarz type action of the type IIB superstring 
in AdS${}_5$ $\times$ S${}^5$ as a sigma model with a coset target 
space PSU($2,2|4$)/[SO(4,1) $\times$ SO(5)]. Then the light-cone 
gauge-fixing of $\kappa$ transformations and reparametrizations 
on the worldsheet were discussed in 
refs.\ \cite{Metsaev:2000yf,Metsaev:2000yu}. 
\par
In this paper we discuss a global symmetry of the type IIB 
superstring in AdS${}_5$ $\times$ S${}^5$ by using a group 
theoretical method. 
The symmetry is represented by the supergroup PSU($2,2|4$). 
We use the worldsheet action of ref.\ \cite{Metsaev:2000yf}, 
where the $\kappa$ symmetry is fixed by the light-cone gauge. 
We obtain explicit forms of the transformation laws for 
the symmetry PSU($2,2|4$) in the light-cone gauge. 
The transformation laws we obtain will be useful in constructing 
the Noether charges for this symmetry \cite{Metsaev:2000yu}. 
They are also useful in finding consistent truncations of the theory, 
which are needed in some recent investigations of the gauge/string 
correspondence 
\cite{Beisert:2004ry,Tseytlin:2004xa,Alday:2005gi,Alday:2005jm}. 
\par
%
%
\newsection{IIB superstring in AdS${}_5$ $\times$ S${}^5$}
The type IIB superstring in AdS${}_5$ $\times$ S${}^5$ can be 
described \cite{Metsaev:1998it} as a sigma model with a target space 
PSU($2,2|4$)/[SO(4,1) $\times$ SO(5)]. 
The supergroup PSU($2,2|4$) contains a bosonic subgroup 
SO(4,2) $\times$ SO(6), which is the isometry of 
AdS${}_5$ $\times$ S${}^5$. Its generators are 
\begin{equation}
T^{\hat{I}} = P^a, J^{ab}, D, K^a, J^i{}_j, Q^{\pm i}, S^{\pm i}, 
\end{equation}
where $P^a, J^{ab}, D, K^a$ are SO(4,2) generators, $J^i{}_j$ 
are SU(4) $\sim$ SO(6) generators, and 
$Q^{\pm i}, S^{\pm i}$ are fermionic generators. 
Here, $a, b, \cdots = 0,1,2,3$ and $i, j, \cdots = 1, 2, 3, 4$ 
denote SO(3,1) and SU(4) indices. 
The (anti-)commutation relations of these generators are given 
in ref.\ \cite{Metsaev:2000yf}, 
whose conventions we use throughout this paper. 
The generators of the subalgebra SO(4,1) $\times$ SO(5) are 
\begin{equation}
J^{ab}, \quad 
\hat{J}^{4a} = K^a + \frac{1}{2} P^a, \quad
J^{A'B'} = - \frac{1}{2} (\gamma^{A'B'})^j{}_i J^i{}_j, 
\end{equation}
where $A', B'=1,2,3,4,5$ are SO(5) indices and $\gamma^{A'}$ 
are SO(5) gamma matrices. 
We use the light-cone coordinates 
$x^\pm = {1 \over \sqrt{2}}(x^3 \pm x^0)$, 
$x = {1 \over \sqrt{2}}(x^1 + i x^2)$, 
$\bar{x} = {1 \over \sqrt{2}}(x^1 - i x^2)$ 
and define $P = P^x$, $\bar{P} = P^{\bar{x}}$, 
$K = K^x$, $\bar{K} = K^{\bar{x}}$. 
\par
We choose a representative of the coset space 
PSU($2,2|4$)/[SO(4,1) $\times$ SO(5)] as 
\begin{eqnarray}
G \A = \A \exp({x^a P^a}) \exp({\theta^{-i}Q^+_i + \theta^-_i Q^{+i} 
+ \theta^{+i} Q^-_i + \theta^+_i Q^{-i}}) \nonu
\A\A \times \exp({\eta^{-i} S^+_i + \eta^-_i S^{+i} 
+ \eta^{+i} S^-_i + \eta^+_i S^{-i}}) 
\exp({\phi D}) \nonu
\A\A \times 
\exp\left({{1 \over 2} i y^{A'} 
(\gamma^{A'})^i{}_j J^j{}_i}\right), 
\label{representative}
\end{eqnarray}
where $\theta^\pm_i = (\theta^{\pm i})^\dagger$, 
$\eta^\pm_i = (\eta^{\pm i})^\dagger$, 
$Q^\pm_i = (Q^{\pm i})^\dagger$, 
$S^\pm_i = (S^{\pm i})^\dagger$. 
The variables $x^a$, $\phi$, $y^{A'}$, $\theta^{\pm i}$, 
$\eta^{\pm i}$ are coordinates of the coset space. We then fix 
the $\kappa$ symmetry by the light-cone gauge condition 
\cite{Metsaev:2000yf} 
\begin{equation}
\theta^{+i} = \eta^{+i} = 0 
\label{gaugecondition}
\end{equation}
and put $\theta^{-i} = \theta^i$, $\eta^{-i} = \eta^i$ for simplicity. 
The left-invariant Cartan one-forms $L^{\hat{I}}$ are defined by 
\begin{eqnarray}
G^{-1} d G \A = \A L^{\hat{I}} T^{\hat{I}} \nonu
\A = \A L_P{}^a P^a + {1 \over 2} L^{ab} J^{ab}
+ L_D D + L_K{}^a K^a + L^j{}_i J^i{}_j 
+ L_Q^{-i} Q^+_i + L_{Qi}^- Q^{+i} \nonu
\A\A + L_Q^{+i} Q^-_i + L_{Qi}^+ Q^{-i}
+ L_S^{-i} S^+_i + L_{Si}^- S^{+i}
+ L_S^{+i} S^-_i + L_{Si}^+ S^{-i}. 
\end{eqnarray}
Using the explicit forms of the Cartan one-forms the world-sheet 
action in the light-cone gauge was obtained in 
ref.\ \cite{Metsaev:2000yf}. 
\par
%
%
\newsection{PSU($2,2|4$) transformations}
According to the general theory of the nonlinear realization 
\cite{Coleman:1969sm,Callan:1969sn} the PSU($2,2|4$) 
transformation of the representative (\ref{representative}) is 
\begin{eqnarray}
G \rightarrow G' = g G h^{-1}(g), 
\end{eqnarray}
where $g$ is an arbitrary element of PSU($2,2|4$), and $h(g)$ is 
a compensating SO(4,1) $\times$ SO(5) transformation which is 
chosen such that $G'$ has a form in eq.\ (\ref{representative}). 
After the light-cone gauge fixing of the $\kappa$ symmetry 
(\ref{gaugecondition}) 
we also need a compensating $\kappa$ transformation. 
An infinitesimal PSU($2,2|4$) transformation is thus written as
\begin{equation}
G^{-1} \delta G 
= G^{-1} \epsilon G - \sigma(\epsilon) + G^{-1} \delta_\kappa G, 
\label{lgtransformation}
\end{equation}
where $\epsilon$ is an arbitrary element of the PSU($2,2|4$) algebra 
\begin{eqnarray}
\epsilon \A = \A \xi^a P^a + \frac{1}{2} \lambda^{ab} J^{ab} 
  + \Lambda D + \zeta^a K^a 
  + v^j{}_i J^i{}_j
  + \epsilon^{-i} Q^+_i + \epsilon^-_i Q^{+i} \nonu 
\A\A + \epsilon^{+i} Q^-_i + \epsilon^+_i Q^{-i} 
  + \beta^{-i}S^+_i + \beta^-_i S^{+i} 
  + \beta^{+i}S^-_i + \beta^+_i S^{-i} 
\end{eqnarray}
and $\sigma(\epsilon)$ is a compensating SO(4,1) $\times$ SO(5) 
transformation 
\begin{equation}
\sigma(\epsilon) 
= \frac{1}{2}\tilde{\lambda}^{ab} J^{ab} 
+ \tilde{\xi}^a \hat{J}^{4a} 
+ \frac{1}{2}\tilde{v}^{A'B'} J^{A'B'}. 
\end{equation}
The last term in eq.~(\ref{lgtransformation}) is a compensating 
$\kappa$ transformation. 
The parameters $\tilde{\lambda}^{ab}$, $\tilde{\xi}^a$, 
$\tilde{v}^{A'B'}$ and those of the $\kappa$ transformation 
depend on $\epsilon$. 
\par
The general $\kappa$ transformation has a form \cite{Metsaev:1998it} 
\begin{eqnarray}
 G^{-1} \delta_\kappa G 
  \A = \A \tilde{\kappa}_Q^{-i} Q_i^+ + \tilde{\kappa}_{Qi}^- Q^{+i} 
  + \tilde{\kappa}_Q^{+i} Q_i^- + \tilde{\kappa}_{Qi}^+ Q^{-i} \nonu
  \A \A + \tilde{\kappa}_S^{-i} S_i^+ + \tilde{\kappa}_{Si}^- S^{+i} 
  + \tilde{\kappa}_S^{+i} S_i^- + + \tilde{\kappa}_{Si}^+ S^{-i} 
  + (\mbox{$J^{ab}$, $J^{A'B'}$ terms}). 
\end{eqnarray}
The coefficients in the present convention are given by 
\begin{eqnarray}
\tilde{\kappa}_Q^{+i} 
\A = \A 2i \left[ \hat{L}_\mu{}^+ \kappa_S^{\mu -i} 
+ \hat{L}_\mu{}^{\bar{x}} \kappa_S^{\mu +i}
+ i \hat{L}_\mu{}^4 \kappa_Q^{\mu +i} 
- L_\mu{}^{A'} (\gamma^{A'})^i{}_j \kappa_Q^{\mu +j}
\right], \nonu
\tilde{\kappa}_Q^{-i} 
\A = \A 2i \left[ \hat{L}_\mu{}^- \kappa_S^{\mu +i} 
- \hat{L}_\mu{}^{x} \kappa_S^{\mu -i}
+ i \hat{L}_\mu{}^4 \kappa_Q^{\mu -i} 
- L_\mu{}^{A'} (\gamma^{A'})^i{}_j \kappa_Q^{\mu -j}
\right], \nonu
\tilde{\kappa}_S^{+i} 
\A = \A -2i \left[ 2 \hat{L}_\mu{}^+ \kappa_Q^{\mu -i} 
+ 2 \hat{L}_\mu{}^{x} \kappa_Q^{\mu +i}
+ i \hat{L}_\mu{}^4 \kappa_S^{\mu +i} 
+ L_\mu{}^{A'} (\gamma^{A'})^i{}_j \kappa_S^{\mu +j}
\right], \nonu
\tilde{\kappa}_S^{-i} 
\A = \A -2i \left[ 2 \hat{L}_\mu{}^- \kappa_Q^{\mu +i} 
-2 \hat{L}_\mu{}^{\bar{x}} \kappa_Q^{\mu -i}
+ i \hat{L}_\mu{}^4 \kappa_S^{\mu -i} 
+ L_\mu{}^{A'} (\gamma^{A'})^i{}_j \kappa_S^{\mu -j}
\right], 
\label{kappatrans}
\end{eqnarray}
where $\mu=0,1$ is a world index on the worldsheet and 
$\kappa_Q^{\mu \pm i}$, $\kappa_S^{\mu \pm i}$ on the right-hand 
sides are independent transformation parameters. 
The $\hat{L}_\mu{}^a$, $\hat{L}_\mu{}^4$, $L_\mu{}^{A'}$ 
are the pullbacks of the following one-forms to the worldsheet 
\begin{equation}
\hat{L}^a = L_P^a - {1 \over 2} L_K^a, \qquad
\hat{L}^4 = - L_D, \qquad
L^{A'} = - {1 \over 2} i (\gamma^{A'})^i{}_j L^j{}_i. 
\label{ls}
\end{equation}
\par
For a general variation of the variables 
$X^M = (x^a, \phi, y^{A'}, \theta^i, \eta^i)$ the variation 
of $G$ in eq.\ (\ref{representative}) is given by 
\begin{eqnarray}
G^{-1} \delta G 
\A = \A \delta X^M L_M{}^{\hat{I}} T^{\hat{I}} \nonu
\A = \A e^\phi \delta x^+ P^-
+ e^\phi \left[ \delta x^- 
- \frac{1}{2} i ( \theta^i \delta \theta_i
+ \theta_i \delta \theta^i ) \right] P^+
+ e^\phi \delta x \bar{P} + e^\phi \delta \bar{x} P \nonu
\A\A  + e^{-\phi} \left[
{1 \over 4}(\eta^2)^2 \delta x^+
+ \frac{1}{2} i ( \eta^i \delta \eta_i
+ \eta_i \delta \eta^i ) \right] K^+
+ \delta \phi D \nonu
\A\A  + \left[ (\delta U U^{-1})^i{}_j 
+ i \left( \tilde{\eta}^i \tilde{\eta}_j 
- {1 \over 4}\eta^2 \delta^i_j \right) \delta x^+ \right] J^j{}_i 
+ e^{\frac{1}{2}\phi} \left( \widetilde{\delta \theta}^i 
+ i \tilde{\eta}^i \delta x \right) Q^+_i \nonu
\A\A  + e^{\frac{1}{2}\phi} \left( \widetilde{\delta \theta}_i 
- i \tilde{\eta}_i \delta \bar{x} \right) Q^{+i}
- i e^{\frac{1}{2}\phi} \tilde{\eta}^i \delta x^+ Q^-_i
+ i e^{\frac{1}{2}\phi} \tilde{\eta}_i \delta x^+ Q^{-i} \nonu
\A\A  + e^{-\frac{1}{2}\phi}\left( \widetilde{\delta \eta}^i 
+ \frac{1}{2} i \eta^2 \tilde{\eta}^i \delta x^+ \right) S^+_i
+ e^{-\frac{1}{2}\phi}\left( \widetilde{\delta \eta}_i 
- \frac{1}{2} i \eta^2 \tilde{\eta}_i \delta x^+ \right) S^{+i}, 
\label{gvariation}
\end{eqnarray}
where we have used the explicit forms of the Cartan 
one-form in the light-cone gauge given in 
ref.~\cite{Metsaev:2000yf}. 
$U^i{}_j$ is the SU(4) matrix determined by the coordinates 
$y^{A'}$
\begin{equation}
U = \cos\frac{|y|}{2} + i \gamma^{A'} n^{A'} \sin\frac{|y|}{2}, 
\label{umatrix}
\end{equation}
where $|y|^2 = y^{A'} y^{A'}$, $n^{A'} = y^{A'}/|y|$, 
and $\tilde{\theta}^i = U^i{}_j \theta^j$, 
$\tilde{\theta}_i = \theta_j (U^{-1})^j{}_i$, etc. 
The compensating transformations in eq.~(\ref{lgtransformation}) 
are chosen such that the total transformation 
(\ref{lgtransformation}) has this form. 
\par
We are now ready to obtain explicit forms of the PSU($2,2|4$) 
transformations. 
We first compute the first term in eq.~(\ref{lgtransformation}). 
Useful formulae to do this are listed in Appendix. 
Then, we choose compensating transformations in the second and 
the third terms such that the total transformations take the 
form in eq.~(\ref{gvariation}). Comparing the results of these 
computations and eq.~(\ref{gvariation}) we obtain the PSU($2,2|4$) 
transformations of the variables $X^M$. 
\par
The transformations for $P^a$, $D$, $J^{+-}$, $J^{+x}$, 
$J^{x\bar{x}}$, $J^{A'B'}$ and $Q^+$ do not need compensating 
$\kappa$ transformations and are easy to obtain. 
They were already given in ref.\ \cite{Metsaev:2000yu}. 
We give them here for completeness. 
\par
\noindent
$\bullet$ $P^a$ transformations: 
\begin{equation}
\delta x^a = \xi^a, \qquad
\delta (\mbox{others}) = 0. 
\label{p}
\end{equation}
%
$\bullet$ $D$ transformations: 
\begin{eqnarray}
\delta x^a \A = \A - \Lambda x^a, \quad
\delta \phi = \Lambda, \quad
(U^{-1} \delta U)^i{}_j = 0, \quad
\delta \theta^i = - \frac{1}{2} \Lambda \theta^i, \quad
\delta \eta^i = \frac{1}{2} \Lambda \eta^i. 
\label{d}
\end{eqnarray}
%
$\bullet$ $J^{+-}$ transformations: 
\begin{eqnarray}
\delta x^+ \A = \A - \lambda^{-+} x^+, \quad
\delta x^- = \lambda^{-+} x^-, \quad
\delta x = \delta \phi 
= (U^{-1} \delta U)^i{}_j = 0, \nonu
\delta \theta^i \A = \A {1 \over 2} \lambda^{-+} \theta^i, \quad
\delta \eta^i = {1 \over 2} \lambda^{-+} \eta^i. 
\label{j+-}
\end{eqnarray}
%
$\bullet$ $J^{+x}$ and $J^{+\bar{x}}$ transformations: 
\begin{equation}
\delta x^- = \lambda^{-\bar{x}} x + \lambda^{-x} \bar{x}, \qquad
\delta x = - \lambda^{-x} x^+, \qquad
\delta (\mbox{others}) = 0. 
\label{j+x}
\end{equation}
%
$\bullet$ $J^{x\bar{x}}$ transformations: 
\begin{eqnarray}
\delta x \A = \A - \lambda^{\bar{x}x} x, \quad
\delta \theta^i = - {1 \over 2} \lambda^{\bar{x}x} \theta^i, \quad
\delta \eta^i = {1 \over 2} \lambda^{\bar{x}x} \eta^i, \quad
\delta (\mbox{others}) = 0. 
\label{jxx}
\end{eqnarray}
%
$\bullet$ $J^i{}_j$ transformations: 
\begin{eqnarray}
\delta x^a \A = \A \delta \phi = 0, \qquad
\delta \theta^i = - v^i{}_j \theta^j, \qquad
\delta\eta^i = - v^i{}_j \eta^j, \nonu
( U^{-1} \delta U )^i{}_j \A = \A v^i{}_j 
+ {1 \over 4} \tilde{v}^{A'B'} (U^{-1}\gamma^{A'B'}U)^i{}_j. 
\label{jij}
\end{eqnarray}
%
$\bullet$ $Q^{+i}$ and $Q^+_i$ transformations: 
\begin{equation}
\delta x^- = \frac{1}{2} i \epsilon^-_i \theta^i 
+ \frac{1}{2} i \epsilon^{-i} \theta_i, \qquad
\delta \theta^i = \epsilon^{-i}, \qquad
\delta (\mbox{others}) = 0. 
\label{q+}
\end{equation}
The transformations for $K^+$ do not need a compensating 
$\kappa$ transformation either. 
\par
\noindent
$\bullet$ $K^+$ transformations: 
\begin{eqnarray}
\delta x^a \A = \A \zeta^- \left( x^+ x^a 
- \frac{1}{2} x \cdot x \, \eta^{a+} 
- \frac{1}{2} e^{-2\phi} \eta^{a+}\right), \qquad
\delta \phi = - \zeta^- x^+, \nonu
\delta \eta^i \A = \A - \zeta^- x^+ \eta^i, \qquad
\delta (\mbox{others}) = 0. 
\label{k+}
\end{eqnarray}
The compensating SO(5) transformation with the parameter 
$\tilde{v}^{A'B'}$ in eq.~(\ref{jij}) is not yet fixed. 
We will determine it and obtain the PSU($2,2|4$) transformation 
of the independent variables $y^{A'}$ in sect.~4. 
\par
Other transformations need compensating $\kappa$ transformations 
and are more involved. 
\par
\noindent
$\bullet$ $J^{-x}$ and $J^{-\bar{x}}$ transformations: 
\begin{eqnarray}
\delta x^+ \A = \A \lambda^{+\bar{x}} x + \lambda^{+x} \bar{x}, \nonu
\delta x^- \A = \A {1 \over 4} i e^{-{3 \over 2}\phi} 
(\eta^i \hat{\kappa}_{Si}^- 
+ \eta_i \hat{\kappa}_S^{-i} ) 
+ {1 \over 2} i e^{-{1 \over 2}\phi} 
(\theta^i \hat{\kappa}_{Qi}^- 
+ \theta_i \hat{\kappa}_Q^{-i} ) \nonu
\A\A - {1 \over 4} i e^{-2\phi} ( 
\lambda^{+x} \theta_i \eta^i 
+ \lambda^{+\bar{x}} \theta^i \eta_i ) \eta^2, \nonu
\delta x \A = \A - \lambda^{+x} \left( x^- - {1 \over 2} i \theta^2 
+ {1 \over 4} i e^{-2\phi} \eta^2 \right), \nonu
\delta \phi \A = \A - {1 \over 2} ( \lambda^{+\bar{x}} \theta^i \eta_i 
- \lambda^{+x} \theta_i \eta^i ), \nonu
(U^{-1} \delta U)^i{}_j 
\A = \A \lambda^{+\bar{x}} \theta^i \eta_j 
+ \lambda^{+x} \theta_j \eta^i 
- (\mbox{trace part})
+ {1 \over 4} \tilde{v}^{A'B'} (U^{-1} \gamma^{A'B'} U)^i{}_j, \nonu
\delta \theta^i 
\A = \A e^{-{1 \over 2}\phi} \hat{\kappa}_Q^{-i} 
- {1 \over 4} \lambda^{+x} e^{-2\phi} \eta^2 \eta^i, \nonu
\delta \eta^i \A = \A e^{{1 \over 2}\phi} 
\hat{\kappa}_S^{-i} 
+ {1 \over 2} \lambda^{+\bar{x}} \eta^2 \theta^i 
+ \lambda^{+x} \theta_j \eta^j \eta^i, 
\label{j-x}
\end{eqnarray}
where we have defined 
\begin{equation}
\hat{\kappa}_Q^{\pm i} = (U^{-1})^i{}_j \tilde{\kappa}_Q^{\pm j}, 
\qquad
\hat{\kappa}_S^{\pm i} = (U^{-1})^i{}_j \tilde{\kappa}_S^{\pm j}. 
\end{equation}
To obtain the form (\ref{gvariation}) we need to choose 
the parameters of the $\kappa$ transformation as 
\begin{equation}
\hat{\kappa}_Q^{+i} 
= \lambda^{+\bar{x}} e^{{1 \over 2}\phi} \theta^i, \qquad
\hat{\kappa}_S^{+i} 
= \lambda^{+x} e^{-{1 \over 2}\phi} \eta^i. 
\label{j-xkappa+}
\end{equation}
As we will see in sect.~4 $\hat{\kappa}_Q^{-i}$, 
$\hat{\kappa}_S^{-i}$ in the transformation (\ref{j-x}) are 
determined from these conditions. 
Similarly, we obtain other transformations and conditions on the 
parameters of the $\kappa$ transformations as follows. 

\noindent
$\bullet$ $K$ and $\bar{K}$ transformations: 
\begin{eqnarray}
\delta x^+ \A = \A ( \bar{\zeta} x + \zeta \bar{x} ) x^+, \nonu
\delta x^- \A = \A 
{1 \over 4} i e^{-{3 \over 2}\phi} ( 
\eta^i \hat{\kappa}_{Si}^- 
+ \eta_i \hat{\kappa}_S^{-i} )
+ {1 \over 2} i e^{-{1 \over 2}\phi} ( 
\theta^i \hat{\kappa}_{Qi}^- 
+ \theta_i \hat{\kappa}_Q^{-i} ) \nonu
\A\A + \bar{\zeta} x \left( x^- + {1 \over 2}i\theta^2 \right) 
+ \zeta \bar{x} \left( x^- - {1 \over 2}i\theta^2 \right) 
+ {1 \over 4} i e^{-2\phi} x^+ ( \bar{\zeta} \theta^i \eta_i 
- \zeta \theta_i \eta^i ) \eta^2, \nonu
\delta x \A = \A \bar{\zeta} x^2 - \zeta x^+ \left( 
x^- - {1 \over 2} i \theta^2 \right) 
- {1 \over 2} \zeta e^{-2\phi} \left( 1 
+ {1 \over 2} i x^+ \eta^2 \right), \nonu
\delta \phi \A = \A - ( \bar{\zeta} x + \zeta \bar{x} ) 
- {1 \over 2} x^+ ( \bar{\zeta} \theta^i \eta_i 
- \zeta \theta_i \eta^i ), \nonu
(U^{-1} \delta U)^i{}_j 
\A = \A x^+ ( \bar{\zeta} \, \theta^i \eta_j 
+ \zeta \, \theta_j \eta^i ) 
- (\mbox{trace part}) 
+ {1 \over 4} \tilde{v}^{A'B'} (U^{-1} \gamma^{A'B'} U)^i{}_j, \nonu
\delta \theta^i 
\A = \A e^{-{1 \over 2}\phi} \hat{\kappa}_Q^{-i}
+ ( \bar{\zeta} x + \zeta \bar{x} ) \theta^i 
+ {1 \over 2} i \zeta e^{-2\phi} \left( 
1 + {1 \over 2}ix^+\eta^2 \right) \eta^i, \nonu
\delta \eta^i 
\A = \A e^{{1 \over 2}\phi} \hat{\kappa}_S^{-i} 
+ i \bar{\zeta} \left( 1 - {1 \over 2}ix^+\eta^2 \right) \theta^i 
- \bar{\zeta} x \eta^i 
+ \zeta x^+ \theta_j \eta^j \eta^i, 
\label{k}
\end{eqnarray}
\begin{equation}
\hat{\kappa}_Q^{+i} 
= e^{{1 \over 2}\phi} \bar{\zeta} x^+ \theta^i, \qquad
\hat{\kappa}_S^{+i} 
= e^{-{1 \over 2}\phi} \zeta x^+ \eta^i. 
\label{kkappa+}
\end{equation}
%
$\bullet$ $K^-$ transformations: 
\begin{eqnarray}
\delta x^+ \A = \A \zeta^+ \left( x^- x^+ - {1 \over 2} x \cdot x 
- {1 \over 2} e^{-2\phi} \right), \nonu
\delta x^- 
\A = \A \zeta^+ \biggl[ {1 \over 2} i e^{-{1 \over 2}\phi} ( 
\theta^i \hat{\kappa}_{Qi}^- 
+ \theta_i \hat{\kappa}_Q^{-i} )
+ {1 \over 4} i e^{-{3 \over 2}\phi} ( 
\eta^i \hat{\kappa}_{Si}^- 
+ \eta_i \hat{\kappa}_S^{-i} ) 
+ (x^-)^2 - {1 \over 4} (\theta^2)^2 \nonu
\A\A 
+ {1 \over 2} e^{-2\phi} \theta^i \eta_i \theta_j \eta^j 
+ {1 \over 4}i e^{-2\phi} \eta^2 ( x \theta_i \eta^i 
+ \bar{x} \theta^i \eta_i ) 
+ {1 \over 16} e^{-4\phi} (\eta^2)^2 \biggr], \nonu
\delta x \A = \A \zeta^+ \left[ \left( 
x^- - {1 \over 2} i \theta^2 \right) x
+ {1 \over 2} e^{-2\phi} \left( \theta^i \eta_i 
+ {1 \over 2} i x \eta^2 \right) \right], \nonu
\delta \phi \A = \A \zeta^+ \left( - x^- 
- {1 \over 2} x \theta_i \eta^i
+ {1 \over 2} \bar{x} \theta^i \eta_i \right), \nonu
(U^{-1} \delta U)^i{}_j 
\A = \A - i \zeta^+ \left( \theta^i \theta_j 
+ i x \eta^i \theta_j 
- i \bar{x} \theta^i \eta_j 
- {1 \over 2} e^{-2\phi} \eta^i \eta_j \right) 
- (\mbox{trace part}) \nonu
\A \A + {1 \over 4} \tilde{v}^{A'B'} 
(U^{-1} \gamma^{A'B'} U)^i{}_j, \nonu
\delta\theta^i 
\A = \A \zeta^+ \left[ e^{-{1 \over 2}\phi} \hat{\kappa}_Q^{-i} 
+ \left( x^- - {1 \over 2} i \theta^2 \right) \theta^i 
- {1 \over 2} i e^{-2\phi} \left( \theta^j \eta_j 
+ {1 \over 2} i x \eta^2 \right) \eta^i \right], \nonu
\delta\eta^i 
\A = \A \zeta^+ \left[ e^{{1 \over 2}\phi} \hat{\kappa}_S^{-i}
+ i \theta_j \eta^j ( \theta^i + i x \eta^i )
- {1 \over 2} \bar{x} \eta^2 \theta^i 
+ {1 \over 4} i e^{-2\phi} \eta^2 \eta^i \right], 
\label{k-}
\end{eqnarray}
\begin{equation}
\hat{\kappa}_Q^{+i} 
= \zeta^+ \left( 
- \bar{x} e^{{1 \over 2}\phi} \theta^i 
+ \frac{1}{2} i e^{-\frac{3}{2}\phi} \eta^i \right), \qquad
\hat{\kappa}_S^{+i} 
= \zeta^+ e^{-{1 \over 2}\phi} i \left( 
\theta^i + i x \eta^i \right). 
\label{k-kappa+}
\end{equation}
%
$\bullet$ $Q^{-i}$ and $Q^-_i$ transformations: 
\begin{eqnarray}
\delta x^+ \A = \A 0, \qquad
\delta x = i \epsilon^+_i \theta^i, \qquad
\delta \phi = - \frac{1}{2}( \epsilon^+_i \eta^i 
- \epsilon^{+i} \eta_i ), \nonu
\delta x^- \A = \A 
  \frac{1}{2}ie^{-\frac{1}{2}\phi}
  (\theta^i \hat{\kappa}_Q^-{}_i 
  + \theta_i \hat{\kappa}_Q^{-i})
  + \frac{1}{4} i e^{-\frac{3}{2}\phi}
  (\eta^i \hat{\kappa}_S^-{}_i 
  + \eta_i \hat{\kappa}_S^{-i}) \nonu
\A\A + \frac{1}{8}i e^{-2\phi} \eta^2
  ( \epsilon^+_i \eta^i 
  + \epsilon^{+i} \eta_i ), \nonu
(U^{-1} \delta U)^i{}_j 
\A = \A - (\epsilon^+_j \eta^i + \epsilon^{+i} \eta_j) 
- (\mbox{trace part})
+ {1 \over 4} \tilde{v}^{A'B'} (U^{-1} \gamma^{A'B'} U)^i{}_j, \nonu
\delta \theta^i \A = \A e^{-\frac{1}{2} \phi} 
  \hat{\kappa}_Q^{-i}, \qquad
\delta \eta^i 
= e^{\frac{1}{2}\phi} \hat{\kappa}_S^{-i}
  - \epsilon^+{}_j \eta^j \eta^i 
  - \frac{1}{2} \eta^2 \epsilon^{+i}, 
\label{q-}
\end{eqnarray}
\begin{equation}
\hat{\kappa}_Q^{+i} 
= - e^{\frac{1}{2}\phi} \epsilon^{+i}, \qquad
\hat{\kappa}_S^{+i} = 0. 
\label{q-kappa+}
\end{equation}
%
$\bullet$ $S^{+i}$ and $S^+_i$ transformations: 
\begin{eqnarray}
\delta x^+ \A = \A 0, \qquad 
\delta x = x^+ \beta^-_i \theta^i, \qquad
\delta \phi = \frac{1}{2} i x^+ ( \beta^-_i \eta^i 
+ \beta^{-i} \eta_i ), \nonu
\delta x^- 
\A = \A \frac{1}{2} i \theta^i ( 
e^{-\frac{1}{2}\phi} \hat{\kappa}^-_{Qi} 
- i\bar{x} \beta^-_i )
+ \frac{1}{2} i \theta_i ( 
e^{-\frac{1}{2}\phi} \hat{\kappa}^{-i}_Q 
+ ix \beta^{-i} ) \nonu
\A\A + \frac{1}{4} i e^{-2 \phi} \eta^i \left[ 
e^{\frac{1}{2}\phi} \hat{\kappa}^-_{Si} 
- \left( 1 - \frac{1}{2} i x^+ \eta^2 \right) 
\beta^-_i \right] \nonu
\A\A + \frac{1}{4} i e^{-2 \phi} \eta_i \left[ 
e^{\frac{1}{2}\phi} \hat{\kappa}^{-i}_S 
- \left( 1 + \frac{1}{2} i x^+ \eta^2 \right) 
\beta^{-i} \right], \nonu
(U^{-1} \delta U)^i{}_j \A = \A i x^+ 
  (\beta^-_j \eta^i - \beta^{-i} \eta_j)
- (\mbox{trace part}) 
+ {1 \over 4} \tilde{v}^{A'B'} (U^{-1} \gamma^{A'B'} U)^i{}_j, \nonu
\delta \theta^i 
\A = \A e^{-\frac{1}{2} \phi} \hat{\kappa}_Q^{-i} 
  - i x \beta^{-i}, \nonu
\delta \eta^i \A = \A 
  e^{\frac{1}{2} \phi} \hat{\kappa}_S^{-i}
  + \left( 1 - \frac{1}{2} i x^+ \eta^2 \right) \beta^{-i}
  + i x^+ \beta^-_j \eta^j \eta^i, 
\label{s+}
\end{eqnarray}
\begin{equation}
\hat{\kappa}_Q^{+i} 
= -i x^+ e^{\frac{1}{2} \phi} \beta^{-i}, \qquad
\hat{\kappa}_S^{+i} = 0. 
\label{s+kappa+}
\end{equation}
%
$\bullet$ $S^{-i}$ and $S^-_i$ transformations: 
\begin{eqnarray}
\delta x^+ \A = \A 0, \qquad
\delta x = x \beta^+_i \theta^i 
+ \frac{1}{2} i e^{-2\phi} \beta^{+i} \eta_i, \nonu
\delta x^- 
\A = \A \frac{1}{4} i e^{-\frac{3}{2}\phi} ( 
\eta^i \hat{\kappa}^-_{Si} 
+ \eta_i \hat{\kappa}^{-i}_S ) 
+ \frac{1}{2} i e^{-\frac{1}{2}\phi} ( 
\theta^i \hat{\kappa}^-_{Qi} 
+ \theta_i \hat{\kappa}^{-i}_Q ) \nonu
\A\A 
  + \frac{1}{4} i e^{-2 \phi} \left[ 
  \left( \theta_j - \frac{1}{2} i \bar{x} \eta_j \right) \eta^j 
  \beta^{+i} \eta_i
  - \left( \theta^j + \frac{1}{2} i x \eta^j \right) \eta_j 
  \beta^+{}_i \eta^i \right] \nonu
\A\A 
  + \frac{1}{2} \left( x^- - \frac{1}{2} i \theta^2 \right) 
  \beta^+{}_i \theta^i
  - \frac{1}{2} \left( x^- + \frac{1}{2} i \theta^2 \right) 
  \beta^{+i} \theta_i, \nonu
\delta \phi \A = \A 
  - \frac{1}{2} \beta^+_i ( \theta^i - i x \eta^i )
  + \frac{1}{2} \beta^{+i} ( \theta_i + i \bar{x} \eta_i), \nonu
(U^{-1} \delta U)^i{}_j \A = \A 
  \beta^+_j (\theta^i + i x \eta^i)
  + \beta^{+i} (\theta_j -  i \bar{x} \eta_j) 
- (\mbox{trace part}) 
+ {1 \over 4} \tilde{v}^{A'B'} (U^{-1} \gamma^{A'B'} U)^i{}_j, \nonu
\delta \theta^i 
\A = \A e^{-\frac{1}{2} \phi} \hat{\kappa}_Q^{-i}
  + \beta^+_j \theta^j \theta^i 
  + i \left( x^- - \frac{1}{2} i \theta^2 \right) \beta^{+i}
  + \frac{1}{2} e^{-2\phi} \beta^{+j} \eta_j \eta^i, \nonu
\delta \eta^i \A = \A 
  e^{\frac{1}{2}\phi}\hat{\kappa}_S^{-i}
  + \beta^+{}_j \eta^j (\theta^i + i x \eta^i)
  - \left( \theta_j - \frac{1}{2} i \bar{x} \eta_j \right) 
  \eta^j \beta^{+i}, 
\label{s-}
\end{eqnarray}
\begin{equation}
\hat{\kappa}_Q^{+i} 
= - i \bar{x} e^{\frac{1}{2} \phi} \beta^{+i}, \qquad
\hat{\kappa}_S^{+i} 
= - e^{-\frac{1}{2} \phi} \beta^{+i}.  
\label{s-kappa+}
\end{equation}
\par
%
%
\newsection{Compensating SO(5) and $\kappa$ transformations}
Here we fix the compensating SO(5) and $\kappa$ transformations 
left undetermined above. 
Let us first consider the SO(5) transformations. 
The PSU($2,2|4$) transformations of $U$ obtained in eqs.~(\ref{jij}), 
(\ref{j-x}), (\ref{k}), (\ref{k-}), (\ref{q-}), (\ref{s+}), (\ref{s-}) 
have a form 
\begin{equation}
( U^{-1} \delta U )^i{}_j = v^i{}_j 
+ {1 \over 4} \tilde{v}^{A'B'} ( U^{-1} \gamma^{A'B'} U )^i{}_j, 
\label{utransformation}
\end{equation}
where $v^i{}_j$ is a given function of the variables $X^M$ 
and the transformation parameters, and $\tilde{v}^{A'B'}$ 
represents a compensating SO(5) transformation. 
On the other hand, a variation of the independent 
variables $y^{A'}$ in eq.~(\ref{umatrix}) gives 
\begin{eqnarray}
U^{-1} \delta U 
\A = \A {1 \over 2} i \gamma^{A'} n^{A'} n^{B'} \delta y^{B'} 
+ {1 \over 2} i {\sin |y| \over |y|} \gamma^{A'} 
( \delta^{A'B'} - n^{A'} n^{B'} ) \delta y^{B'} \nonu
\A\A + {1 \over |y|} \sin^2 {|y| \over 2} \gamma^{A'B'} 
n^{A'} \delta y^{B'}. 
\label{uvariation}
\end{eqnarray}
We choose the compensating SO(5) transformations such that 
eq.~(\ref{utransformation}) has the form (\ref{uvariation}). 
Decomposing $v^i{}_j$ in eq.~(\ref{utransformation}) as 
\begin{equation}
v^i{}_j = {1 \over 2} i v^{A'} (\gamma^{A'})^i{}_j 
- {1 \over 4} v^{A'B'} (\gamma^{A'B'})^i{}_j 
\end{equation}
we find that $\tilde{v}^{A'B'}$ and the PSU($2,2|4$) 
transformations of $y^{A'}$ are given by 
\begin{eqnarray}
\tilde{v}^{A'B'} \A = \A v^{A'B'} 
+ 2 \tan {|y| \over 2} n^{[A'} v^{B']}, \nonu
\delta y^{A'} \A = \A v^{A'B'} y^{B'} + \left[ n^{A'} n^{B'} 
+ {|y| \over \tan |y|} ( \delta^{A'B'} - n^{A'} n^{B'} ) 
\right] v^{B'}. 
\end{eqnarray}
\par
Next, we shall obtain $\hat{\kappa}_Q^{-i}$, $\hat{\kappa}_S^{-i}$ 
from the conditions on $\hat{\kappa}_Q^{+i}$, 
$\hat{\kappa}_S^{+i}$ in eqs.~(\ref{j-xkappa+}), (\ref{kkappa+}), 
(\ref{k-kappa+}), (\ref{q-kappa+}), (\ref{s+kappa+}), 
(\ref{s-kappa+}), which we write as 
\begin{equation}
\hat{\kappa}_Q^{+i} = \tau_Q^i, \qquad
\hat{\kappa}_S^{+i} = \tau_S^i. 
\end{equation}
{}From eq.~(\ref{kappatrans}) these conditions are satisfied if we 
choose the independent $\kappa$ transformation parameters as 
\begin{equation}
\kappa_S^{\mu -i} = - {1 \over 4} i 
{\tau_Q^i \over \hat{L}_\mu{}^+}, \qquad
\kappa_Q^{\mu -i} = {1 \over 8} i 
{\tau_S^i \over \hat{L}_\mu{}^+}, \qquad
\kappa_S^{\mu +i} = \kappa_Q^{\mu +i} = 0, 
\end{equation}
where $\mu=+,-$ are indices of the world-sheet light-cone coordinates.  
Substituting these equations into $\hat{\kappa}_Q^{-i}$, 
$\hat{\kappa}_S^{-i}$ in eq.\ (\ref{kappatrans}) we obtain 
\begin{eqnarray}
\hat{\kappa}_Q^{-i} 
\A = \A - {1 \over 2} \left( 
{\partial_+ x \over \partial_+ x^+} 
+ {\partial_- x \over \partial_- x^+} \right) \tau_Q^i 
+ {1 \over 4} i e^{-\phi} \left( 
{\partial_+ \phi \over \partial_+ x^+} 
+ {\partial_- \phi \over \partial_- x^+} \right) \tau_S^i \nonu
\A\A + {1 \over 4} e^{-\phi} \left( 
{L_+{}^{A'} \over \partial_+ x^+} 
+ {L_-{}^{A'} \over \partial_- x^+} \right) 
(\gamma^{A'})^i{}_j  \tau_S^j, \nonu
\hat{\kappa}_S^{-i} 
\A = \A - {1 \over 2} \left( 
{\partial_+ {\bar{x}} \over \partial_+ x^+} 
+ {\partial_- {\bar{x}} \over \partial_- x^+} \right) \tau_S^i 
+ {1 \over 2} i e^{-\phi} \left( 
{\partial_+ \phi \over \partial_+ x^+} 
+ {\partial_- \phi \over \partial_- x^+} \right) \tau_Q^i \nonu
\A\A - {1 \over 2} e^{-\phi} \left( 
{L_+{}^{A'} \over \partial_+ x^+} 
+ {L_-{}^{A'} \over \partial_- x^+} \right) 
(\gamma^{A'})^i{}_j \tau_Q^j, 
\label{kappa-}
\end{eqnarray}
where we have used the explicit forms of eq.~(\ref{ls}) 
given in ref.~\cite{Metsaev:2000yf} 
\begin{eqnarray}
\hat{L}_\mu{}^+ \A = \A e^\phi \partial_\mu x^+, \qquad
\hat{L}_\mu{}^x = e^\phi \partial_\mu x, \qquad
\hat{L}_\mu{}^4 = - \partial_\mu \phi, \nonu
L_\mu{}^{A'} \A = \A - \frac{1}{2} i (\gamma^{A'})^j{}_i \left[ 
( \partial_\mu U U^{-1})^i{}_j 
+ i \left( \tilde{\eta}^i \tilde{\eta}_j 
- \frac{1}{4} \eta^2 \delta^i_j \right) \partial_\mu x^+ \right]. 
\end{eqnarray}
Using these $\hat{\kappa}^-$'s in eqs.~(\ref{j-x}), (\ref{k}), 
(\ref{k-}), (\ref{q-}), (\ref{s+}), (\ref{s-}) 
we obtain explicit transformation laws. 
\par
{}From eq.\ (\ref{q-}) we see that the $Q^-$ transformation 
of $x^+$ vanishes. This means in particular that the commutator 
of two $Q^-$ transformations is zero on $x^+$, which at first 
sight looks inconsistent with the PSU($2,2|4$) algebra
\begin{equation}
\{ Q^{-i}, Q^-_j \} = i P^- \delta^i_j. 
\label{psuq-q-}
\end{equation}
This apparent inconsistency can be resolved as follows. 
Since we have not fixed a gauge for reparametrizations on the 
worldsheet, the commutator algebra closes up to a reparametrization. 
{}From eqs.~(\ref{q-}), (\ref{q-kappa+}), (\ref{kappa-}) the 
commutator of two $Q^-$ transformations on $x$, which should vanish 
according to the PSU($2,2|4$) algebra (\ref{psuq-q-}), becomes 
\begin{equation}
[ \delta_{Q^-}(\epsilon_1^+), \delta_{Q^-}(\epsilon_2^+) ] x 
= ( \xi^+ \partial_+ + \xi^- \partial_- ) x, 
\end{equation}
where 
\begin{equation}
\xi^\pm = {1 \over 2 \, \partial_\pm x^+} \, i ( 
\epsilon_{2i}^+ \epsilon_1^{+i} 
- \epsilon_{1i}^+ \epsilon_2^{+i} ). 
\end{equation}
This is a reparametrization with the parameters $\xi^\pm$. 
As the reparametrization of $x^+$ with these parameters is 
\begin{equation}
( \xi^+ \partial_+ + \xi^- \partial_- ) x^+ 
= i ( \epsilon_{2i}^+ \epsilon_1^{+i} 
- \epsilon_{1i}^+ \epsilon_2^{+i} ),
\end{equation}
the commutator on $x^+$ can be written as 
\begin{equation}
[ \delta_{Q^-}(\epsilon_1^+), \delta_{Q^-}(\epsilon_2^+) ] x^+ 
= -i ( \epsilon_{2i}^+ \epsilon_1^{+i} 
- \epsilon_{1i}^+ \epsilon_2^{+i} )
+ ( \xi^- \partial_- + \xi^+ \partial_+ ) x^+. 
\end{equation}
The first term on the right-hand side is a $P^-$ transformation
of $x^+$ expected from the PSU($2,2|4$) algebra (\ref{psuq-q-}). 
Thus, the algebra (\ref{psuq-q-}) is satisfied up to a 
reparametrization. 
\par

%

\bigskip

%
\noindent 
{\Large \textbf{Appendix}}
\nopagebreak
\medskip

We summarize formulae useful in computing $G^{-1}\epsilon G$. 
{}From the formula 
\begin{eqnarray}
e^A B e^{-A} 
\A = \A B + [A,B] + {1 \over 2!}[A,[A,B]] 
+ {1 \over 3!}[A,[A,[A,B]]]+ \cdots
\end{eqnarray}
we obtain the following identities. 
\begin{eqnarray}
e^{-x \cdot P} J^{ab} e^{x \cdot P} 
\A = \A J^{ab} - x^a P^b + x^b P^a, \nonu
e^{-x \cdot P} D e^{x \cdot P} \A = \A D - x \cdot P, \nonu
e^{-x \cdot P} K^a e^{x \cdot P} 
\A = \A K^a - x^a D + x^b J^{ba} + x^a x \cdot P 
- \frac{1}{2} x \cdot x \, P^a, \nonu
e^{-x \cdot P} S^{+i} e^{x \cdot P} \A = \A S^{+i} - i x^+ Q^{-i} 
+ i \bar{x} Q^{+i}, \nonu
e^{-x \cdot P} S^{-i} e^{x \cdot P} 
\A = \A S^{-i} - i x^- Q^{+i} - i x Q^{-i}, \nonu
e^{-\theta \cdot Q^+} J^{+-} e^{\theta \cdot Q^+} 
\A = \A J^{+-} + \frac{1}{2} \theta \cdot Q^+, \nonu
e^{-\theta \cdot Q^+} J^{x\bar{x}} e^{\theta \cdot Q^+} 
\A = \A J^{x\bar{x}} - \frac{1}{2} (\theta^i Q^+_i - \theta_i Q^{+i} )
-\frac{1}{2} i \theta^2 P^+, \nonu
e^{-\theta \cdot Q^+} J^{-x} e^{\theta \cdot Q^+} 
\A = \A J^{-x} - \theta^i Q^-_i - \frac{1}{2} i \theta^2 P, \nonu
e^{-\theta \cdot Q^+} D e^{\theta \cdot Q^+} 
\A = \A D -\frac{1}{2} \theta \cdot Q^+, \nonu
e^{-\theta \cdot Q^+} K e^{\theta \cdot Q^+} 
\A = \A K + i \theta^i S^+_i + \frac{1}{2} i \theta^2 J^{+x}, \nonu
e^{-\theta \cdot Q^+} K^- e^{\theta \cdot Q^+} 
\A = \A K^- - i (\theta^i S^-_i - \theta_i S^{-i})
+ \frac{1}{2} i \theta^2 J^{x\bar{x}}
- i \theta^j \theta_i J^i{}_j \nonu
\A\A  - \frac{1}{2} i \theta^2 (\theta^i Q^+_i - \theta_i Q^{+i})
+ {1 \over 4} (\theta^2)^2 P^+, \nonu
e^{-\eta \cdot Q^+} J^i{}_j e^{\eta \cdot Q^+}
\A = \A J^i{}_j - \theta^i Q^+_j + \theta_j Q^{+i}
- i \theta^i \theta_j P^+ - (\mbox{trace part}), \nonu
e^{-\theta \cdot Q^+} Q^{+i} e^{\theta \cdot Q^+} 
\A = \A Q^{+i} + i \theta^i P^+, \nonu
e^{-\theta \cdot Q^+} Q^{-i} e^{\theta \cdot Q^+} 
\A = \A Q^{-i} + i \theta^i \bar{P}, \nonu
e^{-\theta \cdot Q^+} S^{+i} e^{\theta \cdot Q^+} 
\A = \A S^{+i} - \theta^i J^{+\bar{x}}, \nonu
e^{-\theta \cdot Q^+} S^{-i} e^{\theta \cdot Q^+} 
\A = \A S^{-i} + \frac{1}{2} \theta^i ( J^{+-} - J^{x\bar{x}} - D)
+ \theta^j J^i{}_j \nonu
\A\A + \frac{1}{2} \theta^2 Q^{+i} + \theta^i \theta^j Q^+_j
+ \frac{1}{2} i \theta^i \theta^2 P^+, \nonu
e^{-\eta \cdot S^+} P e^{\eta \cdot S^+} 
\A = \A P - i \eta_i Q^{+i} + \frac{1}{2} i \eta^2 J^{+x}, \nonu
e^{- \eta \cdot S^+} P^- e^{\eta \cdot S^+} 
\A = \A P^- - i ( \eta^i Q^-_i - \eta_i Q^{-i})
+ \frac{1}{2} i \eta^2 J^{x\bar{x}}
+ i \eta^i \eta_j J^j{}_i \nonu
\A\A  + \frac{1}{2} i \eta^2 (\eta^i S^+_i - \eta_i S^{+i})
+ {1 \over 4} (\eta^2)^2 K^+, \nonu
e^{- \eta \cdot S^+} D e^{\eta \cdot S^+} 
\A = \A D + \frac{1}{2} \eta \cdot S^+, \nonu
e^{- \eta \cdot S^+} J^{+-} e^{\eta \cdot S^+} 
\A = \A J^{+-} + \frac{1}{2} \eta \cdot S^+, \nonu
e^{- \eta \cdot S^+} J^{x\bar{x}} e^{\eta \cdot S^+} 
\A = \A J^{x\bar{x}} + \frac{1}{2} (\eta^i S^+_i - \eta_i S^{+i})
- \frac{1}{2} i \eta^2 K^+, \nonu
e^{- \eta \cdot S^+} J^{-x} e^{\eta \cdot S^+} 
\A = \A J^{-x} - \eta_i S^{-i} - \frac{1}{2} i \eta^2 K, \nonu
e^{- \eta \cdot S^+} J^i{}_j e^{\eta \cdot S^+}
\A = \A J^i{}_j - \eta^i S^+_j + \eta_j S^{+i} 
+ i \eta^i \eta_j K^+ - (\mbox{trace part}), \nonu
e^{- \eta \cdot S^+} Q^{+i} e^{\eta \cdot S^+} 
\A = \A Q^{+i} + \eta^i J^{+ x}, \nonu
e^{- \eta \cdot S^+} Q^{-i} e^{\eta \cdot S^+} 
\A = \A Q^{-i} - \frac{1}{2} \eta^i ( J^{+-} + J^{x\bar{x}} + D) 
- \eta^j J^i{}_j \nonu
\A\A - \eta^i \eta^j S^+_j - \frac{1}{2} \eta^2 S^{+i}
+ \frac{1}{2} i \eta^i \eta^2 K^+, \nonu
e^{- \eta \cdot S^+} S^{+i} e^{\eta \cdot S^+} 
\A = \A S^{+i} - i \eta^i K^+, \nonu
e^{- \eta \cdot S^+} S^{-i} e^{\eta \cdot S^+} 
\A = \A S^{-i} - i \eta^i K. 
\end{eqnarray}

%

\end{document}